\documentclass{article}
\usepackage{spconf, amsmath,graphicx, float}
\usepackage[flushleft]{threeparttable}
\usepackage{booktabs} % toprule, midrule, bottomrule
\usepackage{url}
\usepackage[symbol]{footmisc}
\title{BIRD-PCC: Bi-directional Range Image-based Deep LiDAR Point Cloud Compression}
\name{Chia-Sheng Liu$^{1}$, Jia-Fong Yeh$^{1}$, Hao Hsu$^{1}$, Hung-Ting Su$^{1}$, Ming-Sui Lee$^{1}$, Winston H. Hsu$^{1,2}$}
\address{$^1$ National Taiwan University \quad $^2$ Mobile Drive Technology}

\begin{document}
%\ninept
%
\maketitle
\begin{abstract}
% Light Detection And Ranging (LiDAR) sensors play a pivotal role in the scenario of autonomous driving. By emitting laser beams to measure the distance between the vehicle and the surrounding objects, the data generated by the LiDAR sensor can provide accurate 3D information. 
% However, the large amount of data collected by LiDAR sensors also brings the issue of data compression. 

The large amount of data collected by LiDAR sensors brings the issue of LiDAR point cloud compression (PCC). Previous works on LiDAR PCC have used range image representations and followed the predictive coding paradigm to create a basic prototype of a coding framework.
% However, their prediction methods could be further improved due to the scheme of unidirectional prediction and the negligence of intrinsic properties of range images. 
However, their prediction methods give an inaccurate result due to the negligence of invalid pixels in range images and the omission of future frames in the time step.
Moreover, their handcrafted design of residual coding methods could not fully exploit spatial redundancy. To remedy this, we propose a coding framework BIRD-PCC. Our prediction module is aware of the coordinates of invalid pixels in range images and takes a bidirectional scheme. Also, we introduce a deep-learned residual coding module that can further exploit spatial redundancy within a residual frame.
Experiments conducted on SemanticKITTI and KITTI-360 datasets show that BIRD-PCC outperforms other methods in most bitrate conditions and generalizes well to unseen environments.
% Experiments conducted on SemanticKITTI dataset demonstrate that BIRD-PCC outperforms other methods in most bitrate conditions.

% Range image-based method for LiDAR data compression is a potential candidate for solution but lacks full exploration, particularly for inter-frame coding. In our work, the problem of range image-based LiDAR sensor data inter-frame compression is investigated. Our inter-frame compression framework follows the predictive coding paradigm and consists of a prediction module and a residual coding module. For the prediction module, we take the previous and next frame in time steps as reference frames, aiming to perform a frame-level prediction of the to-be-encoded frame.
% As for the residual coding module, different from previous methods that first apply quantization followed by a lossless coder, we introduce a learning-based approach that can further exploit spatial redundancy within the residual frame. 
% Experiments conducted on the SemanticKITTI dataset demonstrate that our method outperforms other range image-based methods, especially at low bitrate conditions.

\end{abstract}
\section{Introduction}

\begin{figure}[!t]
\center
\includegraphics[width=0.9\linewidth]{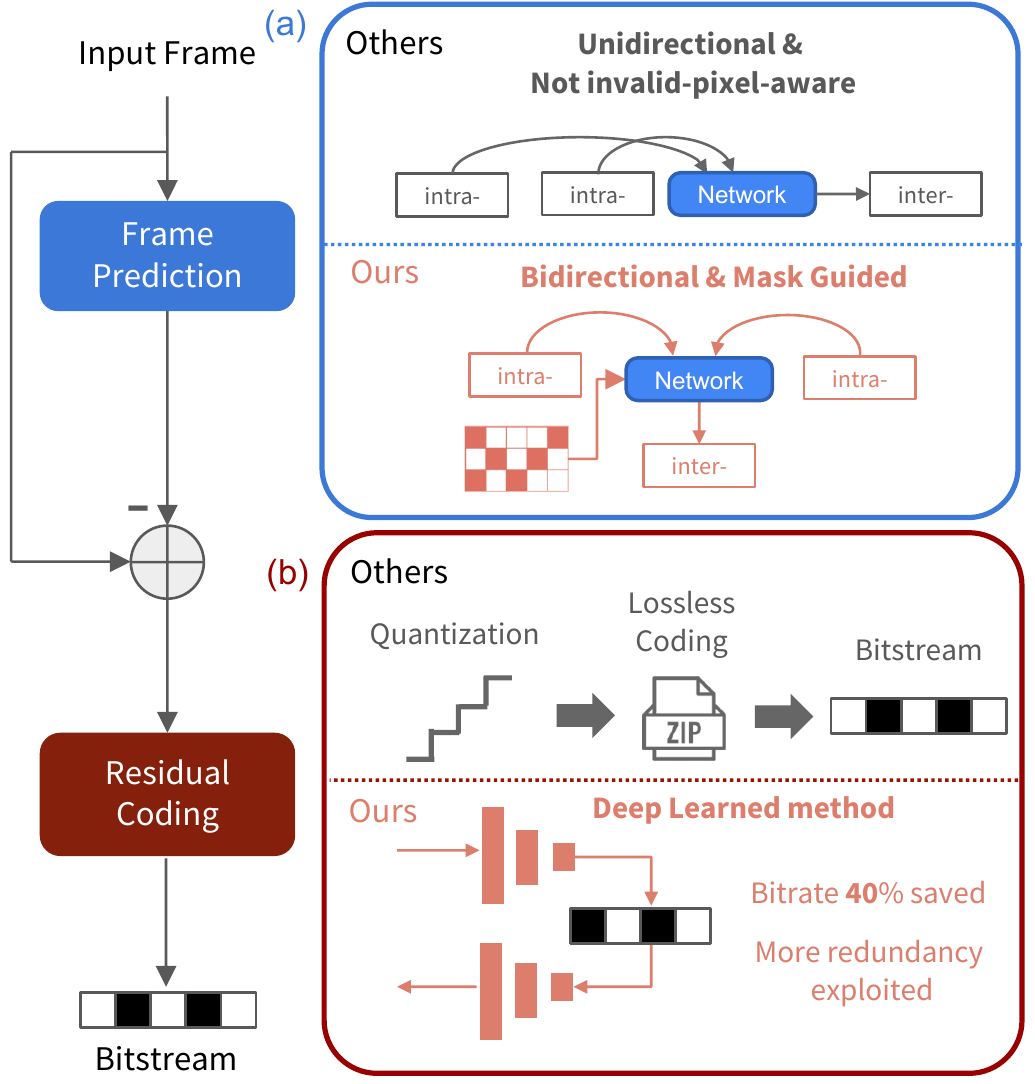}
\vspace{-0.3cm}
\caption{\textbf{Comparison between our method and others.} (a) The prior method omits the coordinates of invalid pixels in range images, resulting in inaccurate predictions (shown in Figure~\ref{prediction_result}) . Also, it only takes the previous frames as reference frames. (b) The prior handcrafted methods are hard to capture complex dependencies, leading to inferior compression ratio.}

\vspace{-0.4cm}
\label{Comparison}
\end{figure}

\label{sec:intro}
Recently, LiDAR point cloud compression (PCC) has been gaining attention in multimedia compression. By providing 3D information, LiDAR point clouds are used to perceive the surrounding environment in the scenario of autonomous driving. However, the storage or transmission of LiDAR data is costly. This is why we delve into the challenging problem of LiDAR point cloud compression in this paper.
% For example, if the Velodyne 64E LiDAR is operated to collect data for 150 seconds, it will generate about 2GB of data. When it comes to a long driving trip, the amount of data generated by LiDAR sensors is very significant. 
% Thus, it is an important issue to compress LiDAR point clouds data.

There are two mainstream approaches for LiDAR PCC. One is tree-based, and the other is range image-based. Tree-based approaches are memory-efficient but suffer from the resolution degradation issue, i.e., a decrease in the number of points in the reconstructed point cloud, as observed in \cite{wang2022rpcc}.
Range image-based approaches utilize the operational mechanism of LiDAR and represent point clouds as range images through spherical projection. 
% Range image is another compact representation for LiDAR point cloud. 
Figure~\ref{Comparison} shows an overview of the recent range image-based approaches that follow the predictive coding paradigm. Sun et al. \cite{sun2019clustering} propose an intra-coding method based on instance-based clustering. The insight of their approach is that the regions belonging to the same object have spatial dependencies. Their clustering algorithm segments a range image into disjoint regions. An average depth value is computed to replace all the pixel values for each region. Finally, the residual values between the ground truth and the average depth value are encoded. Wang et al. follow their work \cite{sun2019clustering} and proposed R-PCC \cite{wang2022rpcc}. To improve the runtime of segmentation, they come up with a region-based clustering method by farthest point sampling. 
% In addition, their method classifies each cluster into different salience levels according to the downstream task requirements. Thus, more reconstruction details would be kept for the more salient regions. 
The above works \cite{wang2022rpcc,sun2019clustering} only consider the spatial redundancy of a single frame, while the temporal redundancy is not leveraged. Subsequently, Sun et al. \cite{sun2020novel} are motivated by the traditional video coding standard and develop a coding architecture that considers both spatial and temporal redundancy. Given a frame sequence, their method first classifies every single frame into intra- or inter-frame. Then, the method in \cite{sun2019clustering} is applied for intra-frame coding. As for inter-frame coding, a neural network consisting of convLSTM cells takes the frames of the previous time step as input, and performs frame-level predictions. Finally, the difference between the ground truth and the prediction is quantized and encoded by an off-the-shelf lossless coder such as bzip2, LZ4, or Deflate.
%The previously mentioned method \cite{sun2019clustering} is applied for intra-frame coding. As for inter-frame coding, a neural network consisting of convLSTM cells takes the frames of the previous time step as input, and performs frame-level predictions. Finally, the difference between the ground truth and the prediction is quantized and encoded by an off-the-shelf lossless coder such as bzip2, LZ4, or Deflate.

% \begin{figure*}[!t]
% \center
% \vspace{-1.3cm}
% \includegraphics[width=0.9\linewidth]{fig/fig.1.pdf}
% \vspace{-0.3cm}
% \caption{Method.}
% \label{method}
% \end{figure*}

The prior works on range image-based method provide a good inspiration, while several aspects could be improved. 
First, the characteristics of range images are different from color images. Therefore, it is not a good practice to apply color image prediction algorithms directly on range images. The question of \textit{how to leverage the properties of range images to develop a better compression algorithm} has still not been sufficiently explored. In this work, our prediction algorithm is aware of the coordinates of invalid pixels in the range images. Second, it is not optimal to perform unidirectional prediction, which only considers the frames in the previous time step. We show that bidirectional prediction could achieve higher accuracy.
% That is, the frames of the previous and the next time step are both considered. 
Third, there is still spatial redundancy existing in a residual frame. Applying quantization and then lossless coding leads to poor rate-distortion performance. As an alternative, we employ a deep-learned framework to deal with residual coding. With more redundancy exploited, better rate-distortion performance could our method achieve. To sum up, our contributions are threefold:
\begin{itemize}

\item We propose BIRD-PCC, a LiDAR point cloud compression framework leveraging the properties of range images and adopting a deep-learned residual coding approach.

\item BIRD-PCC makes an accurate prediction by considering the coordinates of invalid pixels and employing a bidirectional scheme. Also, its residual coding module is more capable of exploring dependencies within residual frames.

\item BIRD-PCC achieves a superior rate-distortion performance in most bitrate conditions and generalizes well to an unseen, larger dataset. At the same distortion level, it costs nearly 20\% less bitrate than the baseline range image-based method.

% \item By taking into account the coordinates of the valid or invalid pixels in the range image, we design a simple yet effective network that performs accurate prediction.
% \item We propose an inter-frame compression framework which considers the previous and the next frames in time step as reference frames to make a frame-level prediction.
% \item We adopte a deep-learned approach for residual frame coding. It is more effective in exploiting redundancy and achieves better rate-distortion performance.
\end{itemize}

\section{Related Work}
\label{sec:format}

{\bf Tree-based LiDAR Point Clouds Compression}. The tree data structure is a memory-efficient representation for sparse point clouds. Draco \cite{draco} and MPEG G-PCC \cite{graziosi2020MPEGoverview} are well-known tree-based methods with KD-tree and octree data structures, respectively, but neither uses deep-learned approaches. Recently, several deep-learned octree compression techniques have emerged. Huang et al. \cite{Huang_2020_CVPR} propose to use stacks of MLPs that extract tree node features and predict the distribution of a node symbol, given the features of the ancestor nodes. To take temporal dependencies into account, Biswas et al. \cite{biswas2020muscle} propose using the node features from the previous time step so that the entropy model can make a more accurate prediction of the distribution. To further exploit the dependencies between tree nodes, Fu et al. \cite{OctAttention} propose an entropy model that considers not only ancestor nodes but also sibling nodes. Besides, they propose an attention mechanism for calculating the weights of different nodes. 

Tree-based methods have been widely explored in recent years and thought of as solutions for LiDAR point cloud compression. However, the number of points of the decoded point cloud is not guaranteed to be the same as that of the original point cloud. This may lead to a divergence of resolution between the ground truth and the decoded point cloud, while range image-based methods do not suffer from this issue.

\section{THE PROPOSED METHOD}
\label{sec:pagestyle}
% This section will first give an overall description of the entire inter-frame coding framework. Then, we will introduce the design of the prediction module and the residual coding module in detail.
\subsection{Overview}
\label{ssec:probForm}
\begin{figure*}[!ht]
\centering
% \vspace{-1cm}
\includegraphics[width=0.9\linewidth, height=5.5cm]{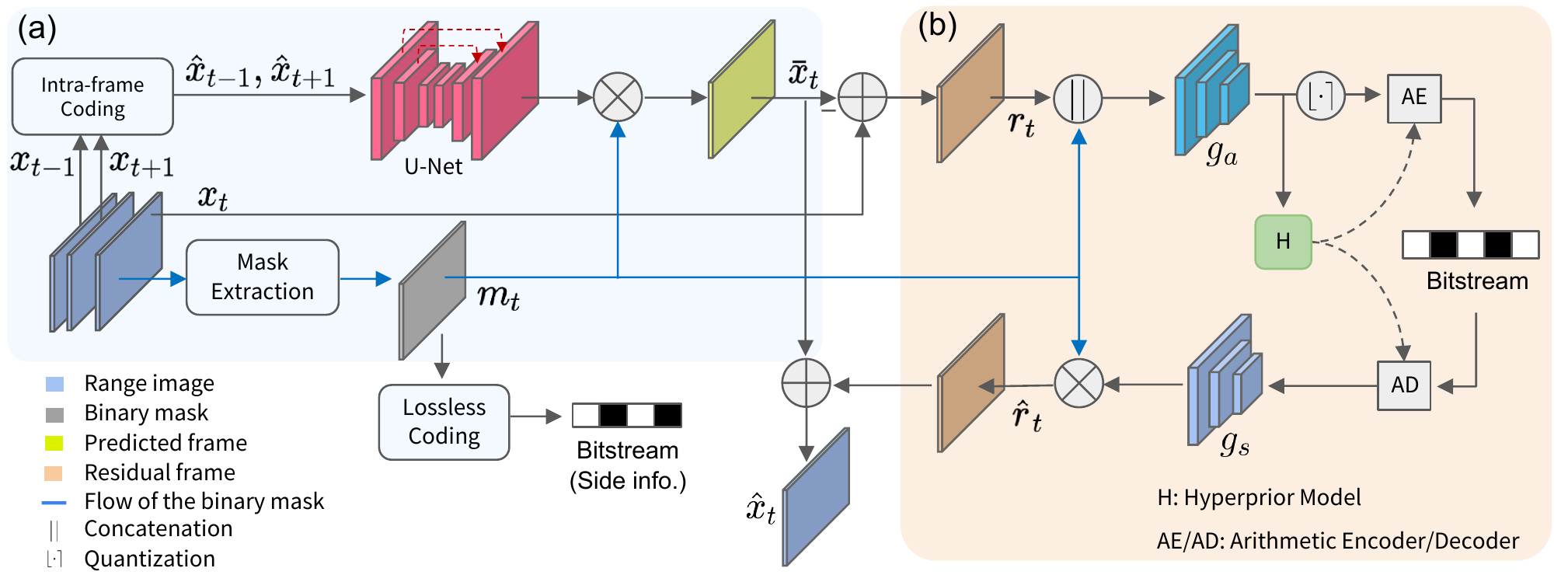}
\vspace{-0.3cm}
\caption{\textbf{Our inter-frame compression pipeline.} (a) The prediction network takes decoded intra-frames $\hat{x}_{t-1}, \hat{x}_{t+1}$ and the mask $m_t$ as inputs and generates a prediction $\bar{x}_t$. (b) The residual coding network $g_a$ and $g_s$ consist of stacks of convolutions, residual blocks  \cite{He_2022_CVPR} and attention blocks \cite{Cheng_2020_CVPR}. For more details about the hyperprior model, please refer to \cite{BalleMSHJ18}. Note that both the encoded mask and the encoded residual frame should be considered as the bitstream to be stored or transmitted.}

\label{pipeline}
\end{figure*}

Our compression pipeline is illustrated in Figure~\ref{pipeline}. 
The setting is consistent with prior works \cite{wang2022rpcc,sun2019clustering,sun2020novel}, focusing solely on 3D geometry coordinates rather than point attributes.
Given a sequence of LiDAR point clouds, each point cloud is converted into a range image representation $\boldsymbol{x} = \left\{ {x_1, x_2, \cdots, x_T}\right\}$ following the formula in section \uppercase\expandafter{\romannumeral 3}. of \cite{wang2022rpcc}. The range image sequence is then split into a bunch of coding units. For each coding unit, the first and last frames in the time steps are treated as intra-frames, while the middle $k$ frames are treated as inter-frames. Note that $k$ is a hyper-parameter, and we take $k=1$ in this work.
Assuming that at the time step $t$, the coding unit is denoted by $ \left\{ {{x}_{t-1}, x_t,{x}_{t+1}}\right\}$.
R-PCC \cite{wang2022rpcc} is adopted to deal with intra-frame coding. As for inter-frame, our prediction module takes decoded intra-frame as reference frames and generates a prediction $\bar{x}$ (section \ref{ssec:prediction}).
The residual frame $r_t$ is formulated as the difference between the ground truth frame $x_t$ and the predicted frame $\bar{x}_t$. Subsequently, the residual frame is encoded into a bitstream through the residual coding module (section \ref{ssec:ResCoding}). 

\subsection{Inter-frame Prediction}
\label{ssec:prediction}

A few research \cite{jiang2018super,zhao2022rangeinet,mersch2021corl} has illustrated that a U-Net architecture with skip connections is capable of frame interpolation or extrapolation.
% When it comes to frame interpolation or extrapolation, U-Net is a widely adopted network architecture\cite{jiang2018super,zhao2022rangeinet,mersch2021corl}. 
% One naive approach is to use the reference frames as the input of an U-Net, and the output is the predicted frame.
A simple approach for frame prediction is to naively set the reference frames and predicted frames as the input and output of the U-Net, respectively. However, this will lead to a low-quality range image output. To address this issue, motion information like optical flow is commonly utilized to make a more accurate prediction in the field of deep-learned video compression \cite{Lu_2019_CVPR,agustsson2020scale}. However, this strategy does not apply to range images due to the intrinsic difference between color images and range images. 

It is noticed that several of the coordinates in the range image are invalid pixels, where the pixel values are zero. These invalid pixels are caused by the absence of LiDAR laser signal return. In our prediction design, we leverage this property and demonstrate that the coordinates of these invalid pixels are essential information to perform an accurate prediction. Precisely, we extract a binary mask $m_t$ from the inter-frame $x_t$ following the formula shown below:
\vspace{-0.12cm}
\begin{align}
 m_t(i,j) = \begin{cases}
0 & \text{ if } x_t(i,j)=0 \\
1 & \text{ if } x_t(i,j)\neq 0 
\end{cases},
\end{align}
where $(i,j)$ is image coordinates. This binary mask is applied as a filter to the last layer of the U-Net, refining the prediction result. Note that the binary mask is considered as side information \cite{cover1999elements}. Thus, we losslessly encode it into a bitstream by bzip2. As for network training, the objective function is the L1 loss between the ground truth and the predicted frame.
% This binary mask is appied as a filter to the last layer of the U-Net, refining the prediction result. Note that the binary mask is considered as side information \cite{cover1999elements}. Thus, we encode it by a lossless coder into bitstream. As for network training, L1 loss betweeen the ground truth and predicted frame is set to be the objective function.

% This mask tells the network which coordinates are valid or invalid. 

\subsection{Residual Frame Coding}
\label{ssec:ResCoding}
% The task of residual coding is to encode the residual frame into a bitstream. The method of prior work quantizes the residual frame and uses a lossless coder to encode the quantized residual frame into a bitstream. We claim that this method can be improved further. Two residual frame coding methods are provided below. One is handcrafted, and the other is deep-learning based.

% \subsubsection{Handcrafted method}
% \label{sssec:Handcrafted_residual_coding}

% We observe that we do not need to encode the entire quantized residual image because there are coordinates where the values can be directly inferred. To be precise, we can determine which coordinates must be invalid pixels from the binary mask generated in the preceding prediction step.
% Because the value of these invalid pixel coordinates must be 0, we do not need to encode the values of these locations. 
% Only the values of the valid pixels need to be encoded. This observation helps us reduce the size of the bitstream.
% \subsubsection{Deep-learned method}
% \label{sssec:DL_residual_coding}

The task of residual frame coding is to encode a residual frame into a bitstream. The method of prior work \cite{wang2022rpcc,sun2019clustering,sun2020novel} first quantizes the residual frame and applies a lossless coder such as bzip2 to perform residual coding.
% We notice that there could be spatial dependency in  quantized residual frames. It is because the prediction in the preceding stage may not be accurate enough. As a consequence, the compression ratio of the prior methods is limited. 
However, the prediction in the former stage may not be accurate enough. Hence, there is spatial redundancy within a residual frame. The prior handcrafted methods are not capable of fully exploiting redundancy, leading to a poor compression ratio.

Our learning-based approach is motivated by the recent research on learned color image compression \cite{ He_2022_CVPR, Cheng_2020_CVPR, BalleMSHJ18, NEURIPS2018}. A variational autoencoder (VAE) is adopted to achieve transform coding. We show that it is also applicable to LiDAR residual signals. Specifically, during the encoding stage, an encoder network $g_a$ performs a non-linear transformation that maps the input from a pixel space into a latent space. To assist in the feature extraction performed by $g_a$, we concatenate the binary mask $m_t$ and the input residual frame $r_t$. The resultant latent codes are then quantized and losslessly encoded into a bitstream by an arithmetic coder \cite{rissanen1979arithmetic}. In order to estimate the distributions of the quantized latent codes, several entropy models have been proposed \cite{Cheng_2020_CVPR, BalleMSHJ18, NEURIPS2018}. We opt for the hyper-prior model \cite{BalleMSHJ18} in our design due to its simplicity and efficiency. At the decoding stage, the decoder network $g_s$ inversely transforms the quantized latent codes into a reconstructed residual frame $\hat{r}_t$. The binary mask $m_t$ is applied as a filter to the last layer of $g_s$, refining the decoded residual frame. The Lagrangian training objective is a combination of rate and distortion (MSE), the same as \cite{BalleMSHJ18}.

\begin{figure*}[!t]
\center
% \vspace{-0.5cm}
\includegraphics[width=0.89\textwidth, height=3.8cm]{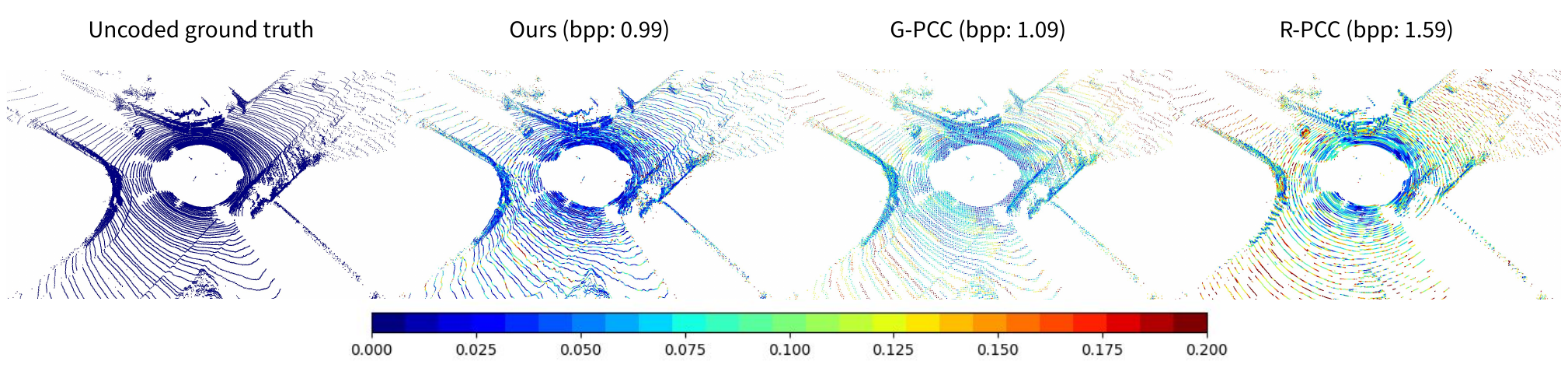}
\vspace{-0.52cm}
\caption{\textbf{Visualization of the ground truth and decoded point clouds of various compression methods.} From left to right: Uncoded ground truth, Ours, G-PCC, and R-PCC. The error color bar is shown at the bottom. Best viewed in color and zoom-in.}
\label{decoded}
\end{figure*}

\begin{figure}[t]
\centering
\vspace{-0.5cm}
\includegraphics[width=0.94\linewidth]{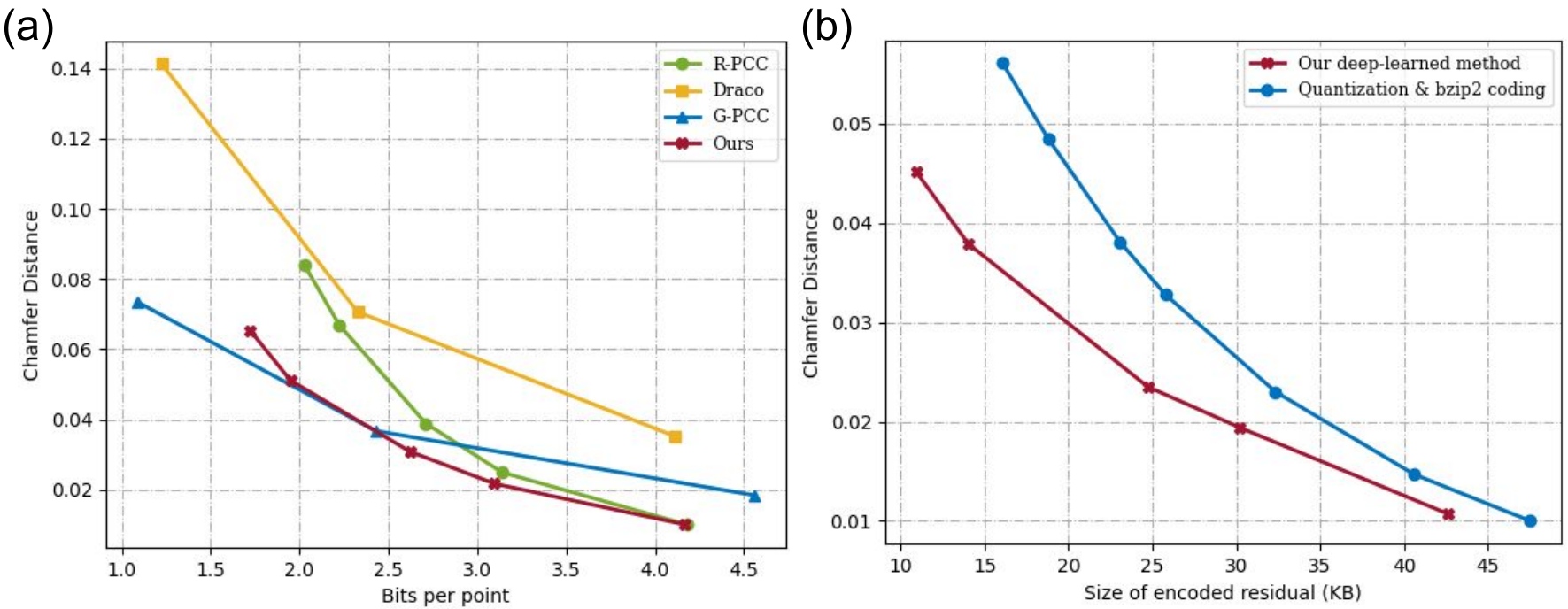} % 
\vspace{-0.45cm}
\caption{\textbf{Rate-distortion (RD) performance.} (a) RD curves with our method and the baselines on KITTI-360. (b) RD curves with various residual coding methods on SemanticKITTI.}
\label{RD}
\end{figure}

\begin{figure}
    \centering
    \vspace{-0.3cm}
    \includegraphics[width=0.92\linewidth,height=3cm]{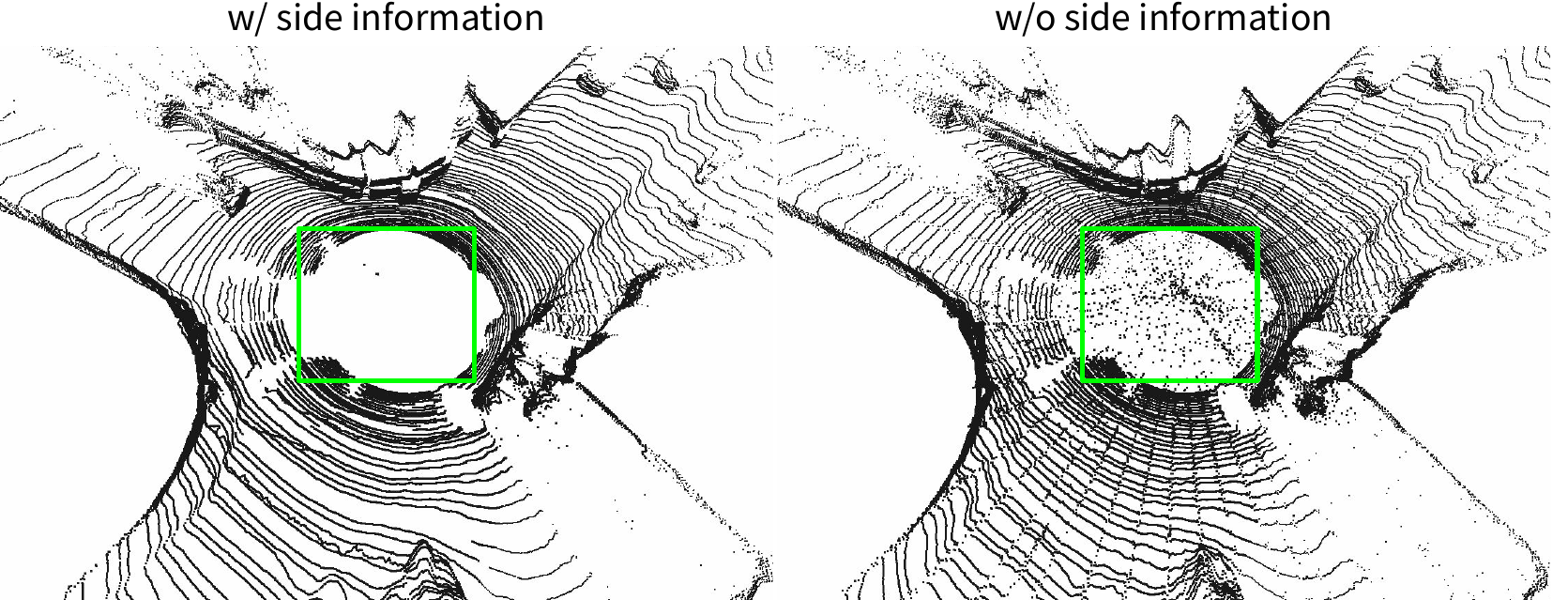} % 
    \vspace{-0.4cm}
    \caption{
    \textbf{The comparison of predicted frames w/ the mask (left) and w/o the mask (right) applied.}
    The predicted frames back-projected to point clouds are shown. It is obvious that the center area bounded by the green region has more noise points in the case of not applying a mask.
    }
    % \vspace{-0.5cm}
    \label{prediction_result}
\end{figure}

\section{EXPERIMENTAL RESULTS}
\label{sec:exp_result}

\subsection{Settings}
\label{ssec:exp_settings}
% {\bf Dataset}. We conduct experiments on SemanticKITTI dataset. The LiDAR data is captured by a Velodyne HDL-64E sensor, which makes the width and height of range images to be (64, 2048). The dataset consists of 22 driving sequences. Our prediction network is trained on sequence 00-10. Due to the need for large amount of training data, we additionaly take sequences 12-20 as the training data for residual coding network. The experiments are tested on sequences 21.
% prediction 18540 scenes
% VAE 34000 scenes
% testing 2721
{\bf Dataset}. 
The prediction network and residual coding network are trained on 18.5k and 34k frames from the SemanticKITTI \cite{behley2019iccv} dataset separately. The evaluations with baselines are conducted on the unseen, larger KITTI-360 \cite{liao2022kitti_360} dataset.

% The networks are trained on SemanticKITTI \cite{behley2019iccv}. Our prediction network is trained on 18.5k frames. Due to the need for large amount of training data, we take 34k frames for residual coding network. The evaluations are conducted on KITTI-360 \cite{liao2022kitti_360}.
% We conduct experiments on SemanticKITTI dataset. Our prediction network is trained on 18.5k frames. Due to the need for large amount of training data, we take 34k frames as the training data for residual coding network. The experiments are tested on 2.7k frames.
\noindent{\bf Evaluation metrics.} Bit-per-point (bpp) is adopted as our bitrate metric. As for the distortion metric, we use Chamfer distance which is also adopted and defined in  \cite{Huang_2020_CVPR,zhou2022riddle}.
% Given a point cloud $P=\left \{ p_1, p_2,...,p_N \right \}$ and its decoded point cloud $\hat{P}=\left \{ \hat{p}_1, \hat{p}_2,...,\hat{p}_M \right \}$, Chamfer distance can be calculated with the following formula: 

% \begin{align}
%  CD(P, \hat{P}) &= \frac{1}{\left |  P\right |}\sum_{i} \min_{j}\left \|  p_i-\hat{p}_j\right \|_2 \\
%  CD_{sym}(P,  \hat{P}) &= \max{\left \{ CD(P, \hat{P}), CD(\hat{P}, P) \right \}}
% \end{align}

\noindent{\bf Baselines.}
Two tree-based baselines and one range image-based baseline are selected in experiments.
Draco \cite{draco} is an open-source compression algorithm based on kd-tree released by Google.
G-PCC \cite{graziosi2020MPEGoverview} is a compression standard based on octree proposed by MPEG. R-PCC \cite{wang2022rpcc} is a range image-based method that only performs intra-frame coding.

% \noindent{\bf Training Loss.}
% The prediction module and residual coding module are trained separately. For the prediction network, L1 loss between ground turth frame and predicted frame is adopted. As for residual coding network, we follow the rate-distortion loss widely used in neural image compression\cite{BalleMSHJ18, NEURIPS2018, Cheng_2020_CVPR, He_2022_CVPR}.

\subsection{Quantitative Results of Rate-distortion Curves}
\label{ssec:RD_Results}
% To show the generalizability of BIRD-PCC, we compare the rate-distortion (RD) performance with the selected baselines on the unseen KITTI-360 dataset. The quantitative results are shown in Figure~\ref{RD}(a). It is obvious that our method outperforms R-PCC at low bitrate conditions and surpasses tree-based methods at high bitrate conditions.
To show the generalizability of BIRD-PCC, we compare the rate-distortion performance with the selected baselines on the unseen KITTI-360 dataset.
Figure~\ref{RD}(a) shows the quantitative results, indicating our method outperforms R-PCC at low bitrates and surpasses tree-based methods at high bitrates.

\subsection{Qualitative Results}
\label{ssec:Visualization}
In Figure~\ref{decoded}, the decoded point clouds of our method, R-PCC, and G-PCC are visualized. Our method could achieve lower distortion while the bitrate is still less than the others.
% Moreover, our method maintains the densities of points, while octree-based G-PCC does not.
Moreover, our method maintains the distribution pattern of points, particularly apparent in the central o-shaped area where the point density should be high. However, octree-based G-PCC fails to maintain the pattern due to voxelization.

\vspace{-0.5cm}

\begin{table}[!h]
\caption{Prediction results of various designs.}
\label{Prediction}
  \begin{threeparttable}[t]
  \centering
  \small
       \begin{tabular}{lrrr}
    \toprule
     Method  & L1 ↓  & RMSE\tnote{b} ↓ & Acc.@0.1\tnote{b} ↑  \\
    \midrule
    Avg. dist. between frames  & 2.250          & 0.322          & 49.2\%    \\
    Bidirectional w/o SI\tnote{a} & 1.354          & 0.393          & 52.5\%    \\
    Unidirectional w/ SI\tnote{a} & 0.338 & 0.333 & 59.7\%  \\
    Bidirectional w/ SI\tnote{a}  & \textbf{0.288} & \textbf{0.302} & \textbf{63.6\%}  \\

     \bottomrule
  \end{tabular}
     \begin{tablenotes}
     \item[a] SI: Side Information (Mask)
     \item[b] The RMSE and accuracy metric is defined in \cite{sun2020novel} and \cite{zhou2022riddle}.
   \end{tablenotes}
    \end{threeparttable}%
  \label{tab:addlabel}%
\end{table}%

\subsection{Ablation Study}
% In this section, we compare different implementations with regard to prediction module and residual frame coding module.
\label{ssec:ablation}

\noindent{\bf Choice of prediction module.} We compare the prediction quality of different choices of frameworks.
In Table~\ref{Prediction}, we show the effect of the binary mask during frame prediction. Also, we compare the unidirectional approach \cite{sun2020novel} and the bidirectional approach. For the sake of fairness, we use the previous two frames as reference frames to make a prediction, and the network backbone is still a U-Net in the experiment of the unidirectional case. The experiments show that our design has the best prediction quality in all metrics. To exhibit the benefit of the binary mask, we visualize the prediction results in Figure~\ref{prediction_result}.

\noindent{\bf Choice of residual coding module.} Different methods of residual frame coding are compared, including our deep-learned method and the prior handcrafted method.
% We fix the same quality of R-PCC decoded frames to perform prediction and get the residual frame. 
The same residual frames are encoded by various methods under different bitrate conditions. The results are shown in Figure \ref{RD}(b). It is worth mentioning that our method could achieve 40\% less bitrate than the prior method at the same level of distortion.

\section{CONCLUSIONS}
\label{sec:conclusion}
With the growth of the amount of data captured by LIDAR sensors, it becomes increasingly worthwhile to explore how to compress this type of data effectively. In this paper, we present BIRD-PCC, aiming to find a solution for inter-frame coding for LiDAR point clouds. We use a bidirectional prediction framework and consider the coordinates of invalid pixels to make a more accurate prediction. 
In addition, we use a learning-based approach to exploit the redundancy in the residual frame more effectively. The experimental results on the KITTI-360 dataset show that the rate-distortion performance of BIRD-PCC outperforms the prior range image-based method and generalizes well to unseen environments.

\section{acknowledgement}
\label{sec:acknowledgement}
This work was supported in part by National Science and Technology Council, Taiwan, under Grant NSTC 111-2634-F-002-022, by Qualcomm through a Taiwan University Research Collaboration Project, and by NOVATEK fellowship. 

% Below is an example of how to insert images. Delete the ``\vspace'' line,
% uncomment the preceding line ``\centerline...'' and replace ``imageX.ps''
% with a suitable PostScript file name.
% -------------------------------------------------------------------------

\clearpage
% To start a new column (but not a new page) and help balance the last-page
% column length use \vfill\pagebreak.
% -------------------------------------------------------------------------
%\vfill
%\pagebreak

\vfill\pagebreak

% References should be produced using the bibtex program from suitable
% BiBTeX files (here: strings, refs, manuals). The IEEEbib.bst bibliography
% style file from IEEE produces unsorted bibliography list.
% -------------------------------------------------------------------------
\bibliographystyle{IEEEbib}
\bibliography{strings,refs}

\end{document}